# Concerning Quadratic Interaction in the Quantum Cheshire Cat Experiment

*W.M. Stuckey*
*Department of Physics, Elizabethtown College, Elizabethtown, PA  17022*
*stuckeym@etown.edu*

*Michael Silberstein*
*Department of Philosophy, Elizabethtown College, Elizabethtown, PA  17022*
*Department of Philosophy, University of Maryland, College Park, MD  20742*
*silbermd@etown.edu*

*Timothy McDevitt*
*Department of Mathematics, Elizabethtown College, Elizabethtown, PA  17022,*
*mcdevittt@etown.edu*

**Abstract**
In a July 2014 *Nature Communications* paper, Denkmayr *et al*. claim to have instantiated the so-called quantum Cheshire Cat experiment using neutron interferometry. Crucial to this claim are the weak values which must imply the quantum Cheshire Cat interpretation, i.e., "the neutron and its spin are spatially separated" in their experiment. While they measured the correct weak values for the quantum Cheshire Cat interpretation, the corresponding implications do not obtain because, as we show, those weak values were measured with both a quadratic and a linear magnetic field $B_z$ interaction. We show explicitly how those weak values imply quantum Cheshire Cat if the $B_z$ interaction is linear and then we show how the quadratic $B_z$ interaction destroys the quantum Cheshire Cat implications of those weak values. Since both linear and quadratic $B_z$ interactions contribute equally to the neutron intensity in this experiment, the deviant weak value implications are unavoidable. Because weak values were used successfully to compute neutron intensities for weak $B_z$ in this experiment, it is clearly the case that one cannot make ontological inferences from weak values without taking into account the corresponding interaction strength.

# 1. INTRODUCTION

Using a neutron interferometer, Denkmayr et al. claim[1] to have instantiated, for the first time, the quantum Cheshire Cat experiment. In a quantum Cheshire Cat experiment[2], a particle is spatially separated from one of its properties, just as the Cheshire Cat can be spatially separated from its grin in the Lewis Carroll story *Alice's Adventures in Wonderland*[3] (see Denkmayr et al.'s Figure 1, included here[1]). Specifically, they claim to have separated neutrons from the *z* component of their spin, i.e., the neutrons take one path through the interferometer while the *z* component of the spin of those neutrons takes the other path through the interferometer. Corrêa et al. showed[4] that the quantum Cheshire Cat experiment in general, and the so-called "qualitative result[2]" of Denkmayr et al. specifically (section 3), can be explained by quantum interference. Of course, that quantum Cheshire Cat can be understood by quantum interference does not make it less interesting, since quantum interference is a pressing issue for foundationalists. While we do not object to the possibility of a quantum Cheshire Cat experiment, we argue herein that the quantum Cheshire Cat interpretation requires *both* the necessary set of weak values (section 4) *and* a linear interaction (section 5). Thus, in general, one cannot make ontological inferences from weak values without taking into account the commensurate interaction strength. We make this argument using the Denkmayr et al. experiment because it contains an unavoidable quadratic contribution from the magnetic field $B_z$ interaction which destroys the quantum Cheshire Cat implications of the weak values (section 5 and Figure 2). That is, in order to instantiate quantum Cheshire Cat in this experiment the weak values (section 4) $\langle \hat{\Pi}_I \rangle_W = 0$ and $\langle \hat{\Pi}_{II} \rangle_W = 1$ must imply that the neutrons at detector O took path II through the interferometer while the weak values $\langle \hat{\sigma}_z \hat{\Pi}_I \rangle_W = 1$ and $\langle \hat{\sigma}_z \hat{\Pi}_{II} \rangle_W = 0$ must imply that the *z* component of the neutrons' spin at detector O took path I through the interferometer. In that case, they would be able to make the following (italicized) inference using the weak values they measured in this experiment:

---

[1] Figures 1 – 4 herein are remixed from the originals in Denkmayr et al. per Creative Commons CC-BY license.
[2] In correspondence with Denkmayr et al., they referred to the variable-χ interference pattern as a "qualitative result." This "qualitative result" does not constitute quantum Cheshire Cat, which must be established quantitatively via weak values. The point we are trying to establish in this paper is that in addition to the weak values, the quantum Cheshire Cat interpretation requires a corresponding linear interaction to remain viable.

The appropriate observable to ascertain the weak value of the neutrons' spin component on path $j$ is $\left\langle \hat{\sigma}_z \hat{\Pi}_j \right\rangle_W$. The computation of the weak values yields $\left\langle \hat{\sigma}_z \hat{\Pi}_I \right\rangle_W = 1$ and $\left\langle \hat{\sigma}_z \hat{\Pi}_{II} \right\rangle_W = 0$. *On average, a weak interaction coupling with a probe on path II does not affect the state of that probe, as if there was effectively no spin component travelling along the path* [our emphasis].

Stated otherwise, "Therefore, any probe system that interacts with the Cheshire Cat system *weakly enough* will on average be affected as if the neutron and [the z component of] its spin are spatially separated [our emphasis]." So, they needed to show that it is possible to introduce a *weak enough* magnetic field $B_z$ to the interferometer path I such that there would be a measureable effect, $\left\langle \hat{\sigma}_z \hat{\Pi}_I \right\rangle_W = 1$, while that same $B_z$ on path II would show no such effect, $\left\langle \hat{\sigma}_z \hat{\Pi}_{II} \right\rangle_W = 0$, i.e., the grin is on the lower path (I) not the upper path (II), as depicted in their Figure 1. As we will show, "weakly enough" means "linearly" and the quadratic contribution results in an observable effect (grin) on the upper path (II) which destroys quantum Cheshire Cat (as depicted by Figure 2). In the corresponding case with photons, for example, Corrêa *et al.* show[5] that in order to get the photon amplitude in the quantum Cheshire Cat experiment one must keep only the linear terms in the expansion of the total amplitude, their Eq (5), to obtain the amplitude for quantum Cheshire Cat, their Eq (7). While we cannot make a direct correspondence between the Denkmayr *et al.* neutron experiment and the Correa *et al.* photon experiment due to the lack of an explicit "probe state" in the neutron version, we argue here that something similar is at work. We will show that it is impossible to make $B_z$ weak enough to decouple observationally from the neutron's magnetic moment on path II without also having it decouple on path I, precisely because the quadratic term in the $B_z$ interaction contributes observationally as much as the linear term. Therefore, we posit that weak values can only be interpreted as meaning the particle and one of its properties have been spatially separated, i.e., quantum Cheshire Cat interpretation, if the relevant interaction can be made weak enough to render the quadratic contribution to the interaction negligible.

Conversely, if the four weak values measured by Denkmayr *et al.*, in and of themselves, constitute quantum Cheshire Cat, the Denkmayr *et al.* experiment is quantum Cheshire Cat, but serves as a reductio against the quantum Cheshire Cat interpretation (because there is an observable effect for $B_z$ in either path). We suspect that the weak values community would rather opt to simply add the qualifier of a linear interaction to the interpretation of the weak values. In that case, Denkmayr *et al*. did not do quantum Cheshire Cat, but the quantum Cheshire Cat interpretation is still viable. Either way, this strikes us an important point for the weak values program in general because Denkmayr *et al*. have established that weak values can be mapped onto experimental outcomes even when there is quadratic interaction. And clearly, the implications of the weak values in that situation are not straightforward. Thus, the analysis in this paper should prompt further discussion as to the ontological inferences one can make from weak values.

We will discuss the Denkmayr *et al*. experimental results in section 3, after we briefly review the experiment in section 2. In section 4, we examine the definitions for the weak values in this experiment, explaining what they entail and do not entail. Section 4 is self-contained, but the interested reader may consult background material on weak values[6]. In section 5, we show that the weak values imply quantum Cheshire Cat with a linear $B_z$ interaction, but that implication is destroyed by the quadratic $B_z$ interaction because it leads to an observational effect (intensity) on path II where we need it to disappear (Figures 1 & 2). Since the quadratic $B_z$ term contributes to the intensity with the same magnitude as the linear term, $B_z$ can never be made weak enough to decouple observationally from the neutron's magnetic moment on path II without also decoupling on path I. Therefore, we conclude (section 6) that even though the weak values in this experiment accurately account for the measured neutron intensities, they do not imply the quantum Cheshire Cat interpretation because of the (unavoidable) quadratic interaction. That means one cannot ignore interaction strength when making ontological inferences from weak values.

## 2. THE EXPERIMENT

The experiment is depicted in Denkmayr *et al*.'s Figures 3 & 4 (included here). To understand the essential elements of the experiment, you need to know that spin rotators create $|S_x +\rangle$ (or $|+\rangle$ for short) on path I and $|S_x -\rangle$ (or $|-\rangle$ for short) on path II (brown boxes in Figure 3) just after the neutrons pass through the first beam splitter (entering from the left in Figure 3). Path I is the "lower path" and path II is the "upper path." The two detectors are O (labeled by $I_o$ in yellow boxes of Figures 3 & 4) and H (labeled by $I_H$ in yellow box of Figure 4). A $|-\rangle$ spin selector immediately precedes the detector O (red box labeled SA) while the entire signal is sampled at H. A phase difference χ between the two paths can be introduced and this is represented by the white bar next to the second beam splitter in Figures 3 & 4. Thus, when a partial (weak) absorber (brown bar in Figure 3) is placed in path I it diminishes the $|+\rangle$ amplitude contributing to the amplitude going to the spin selector. But, the $|-\rangle$ spin selector deletes that effect on the amplitude at O, so there is no change in the intensity at O. However, when the partial absorber is placed in path II it diminishes the $|-\rangle$ amplitude contributing to the amplitude at the spin selector, so this decrease in the amplitude obtains at O giving rise to a slight decrease in the intensity at O. The experimenters therefore conclude that the neutrons reaching O are taking path II, i.e., "a minimally disturbing measurement will find the Cat in the upper beam path … ." This part of the experiment is straightforward and requires no detailed analysis. It is the second part of the experiment, i.e., the introduction of a weak magnetic field $B_z$, that yields the controversial part of the conclusion, i.e., "… while its grin will be found in the lower one."

## 3. EXPERIMENTAL RESULTS

Their claim follows from their weak values as computed using the χ = 0 results (as we discuss in section 4), so let us look carefully at that data. Specifically, again, it is only the introduction of a weak $B_z$ that leads to controversy, so we focus on that. The authors give the χ = 0 reference value for the intensity at O as $I^{REF}$ = 11.25(5) counts per second (cps). The theoretical intensity at O (where, again, only $|-\rangle$ is sampled) when the magnetic field $B_z$ is placed in path II (their Eq (14)) is $I_{II/O}^{MAG} = I^{REF} cos^2\left(\frac{\alpha}{2}\right)$, where α is proportional to the magnetic field intensity. [This is independent of χ since $|-\rangle$ amplitude only exists in the upper path in this case.] For α = 20° (given in their paper) this yields a theoretical prediction of 10.91(5) cps, which agrees with their measured value of $I_{II}^{MAG} = 10.93(6)$ cps. Thus, when $B_z$ is introduced in the upper path (II) with small α (small magnetic field intensity), we see a small decrease in the χ = 0 intensity at O due to some of the $|-\rangle$ amplitude in the upper path being converted to $|+\rangle$ amplitude by the magnetic field there. The decrease is given by (to a first approximation)

$$I_{II/O}^{MAG} \approx I^{REF}\left(1-\left(\frac{\alpha}{2}\right)^2\right) \tag{1}$$

or about a 3% reduction in $I^{REF}$. What we see at H in this case is a sinusoidal oscillation in $\chi$ due to interference between $|+\rangle e^{i\chi/2}$ (created from $|-\rangle$ by $B_z$ in the upper path) and $|+\rangle e^{-i\chi/2}$ from the lower path. This and the other interference results prompt Corrêa *et al.* to write, "the results can be explained as simple quantum interference" although, again, it is the non-interference ($\chi = 0$) component of the experiment that determines the weak values necessary to establish the quantum Cheshire Cat interpretation.

When $B_z$ is introduced in the lower path (I), we have $I_{I/O}^{MAG} = \frac{I^{REF}}{2}(3-\cos(\alpha))$ for $\chi = 0$ (using their Eq (13)), which for $I^{REF}$ = 11.25(5) cps and $\alpha = 20^\circ$ gives a theoretical prediction of 11.59(5) cps in agreement with their measured value of $I_I^{MAG} = 11.57(6)$ cps. So, the reason we see an increase in the intensity at O when $B_z$ is placed in the lower path is because it generates an additional $|-\rangle$ term in the amplitude when operating on $|+\rangle$ and $|-\rangle$ is what is measured at O. This slight increase in $|-\rangle$ created by the magnetic field on path I is added to that from the upper path going to O. The increase is given by (to a first approximation)

$$I_{I/O}^{MAG} \approx I^{REF}\left(1+\left(\frac{\alpha}{2}\right)^2\right) \tag{2}$$

or about 3%. So, the increase in intensity at O for $B_z$ in path I is no more pronounced than the 3% decrease in intensity at O for $B_z$ in path II. Indeed, we will show (section 4) that you cannot eliminate the observational effect of weak $B_z$ in path II (by further weakening $B_z$) without also doing so in path I, so you can never observe the grin on path I alone as required for the quantum Cheshire Cat interpretation. Unlike the previous case, we do have oscillation in $\chi$ at O for this case caused by interference between $|-\rangle e^{i\chi/2}$ in the upper path and $|-\rangle e^{-i\chi/2}$ created from $|+\rangle$ by $B_z$ in the lower path. A similar interference creates an oscillation in $\chi$ at H that is approximately out of phase with that at O. This oscillation in $\chi$ at O (so-called "qualitative result") and H does not establish the quantum Cheshire Cat interpretation, nor does it have anything to do with the weak values necessary to establish the quantum Cheshire Cat interpretation. We are now in position to scrutinize the weak values in Denkmayr *et al*.

## 4. WEAK VALUES

Denkmayr *et al*. use their weak values $\langle \hat{\Pi}_I \rangle_W = 0$ and $\langle \hat{\Pi}_{II} \rangle_W = 1$ to infer that the neutrons at detector O took path II through the interferometer and their weak values $\langle \hat{\sigma}_z \hat{\Pi}_I \rangle_W = 1$ and $\langle \hat{\sigma}_z \hat{\Pi}_{II} \rangle_W = 0$ to infer that the *z* component of the neutrons' spin at detector O took path I through the interferometer, i.e., the quantum Cheshire Cat interpretation. However, the data above and a simple analysis of their Eqs (1) – (3) and (10) reveal that their weak values in this experiment do not support their claim. As it turns out, these weak values are necessary, but not sufficient, for establishing the quantum Cheshire Cat interpretation. As we will show, these weak values must be obtained with a linear $B_z$ interaction in order to establish the quantum Cheshire Cat interpretation, but the (observable) $B_z$ interaction in this experiment (necessarily) contains a quadratic contribution.

As they show in their Eq (10), the weak values appear as expansion coefficients in the weak field approximation to the $\chi = 0$ intensities. That is, their Eq (10),

$$I_j^{MAG} = |\langle \psi_f | \psi_i \rangle|^2 \left[ 1 - \frac{\alpha^2}{4} \langle \hat{\Pi}_j \rangle_W + \frac{\alpha^2}{4} |\langle \hat{\sigma}_z \hat{\Pi}_j \rangle_W|^2 \right]$$

with $|\langle \psi_f | \psi_i \rangle|^2 = I^{REF}$, $\langle \hat{\Pi}_I \rangle_W = 0$, $\langle \hat{\Pi}_{II} \rangle_W = 1$, $\langle \hat{\sigma}_z \hat{\Pi}_I \rangle_W = \pm 1$, and $\langle \hat{\sigma}_z \hat{\Pi}_{II} \rangle_W = 0$ is nothing more than $I_{II/O}^{MAG} \approx I^{REF} \left( 1 - \left( \frac{\alpha}{2} \right)^2 \right)$ and $I_{I/O}^{MAG} \approx I^{REF} \left( 1 + \left( \frac{\alpha}{2} \right)^2 \right)$, Eqs (1) & (2) above. So, we see that the weak values they measure follow tautologically from the exact functional forms for the intensity at O, as long as they use a weak enough $B_z$, where "weak enough" in this context means the measured weak values agree with theory within experimental limits. So, contrary to their claim that "a small magnetic field has on average a significant effect only in path I, while it has none in path II," we see that theory tells us the effect of a weak $B_z$ is as pronounced in path II as it is in path I with respect to the $\chi = 0$ intensities, which are the intensities used to measure the weak values needed for the quantum Cheshire Cat interpretation.

Before exploring the definitions for the weak values in this experiment, we point out that the weak values $\langle \hat{\sigma}_z \hat{\Pi}_I \rangle_W = 1$ and $\langle \hat{\sigma}_z \hat{\Pi}_{II} \rangle_W = 0$ are not in some "weak" way related to $\langle \hat{\sigma}_z \rangle$ for each path. A spin $z$ measurement in either path results in $\langle \hat{\sigma}_z \rangle = 0$, since in path I we have

$$\langle + | \hat{\sigma}_z | + \rangle = -\langle + | - \rangle = 0 \tag{3}$$

while in path II we have

$$\langle - | \hat{\sigma}_z | - \rangle = -\langle - | + \rangle = 0 \tag{4}$$

This follows trivially from the fact that a spin $z$ measurement of either $|S_x +\rangle$ or $|S_x -\rangle$ gives 50% up and 50% down outcomes which average to zero. This result is not a function of $B_z$ field strength, so there is no "weak field approximation" for it, as there is for intensity (their Eq (10), for example). And, obviously, $\langle \hat{\sigma}_z \rangle = 0$ does not mean that "*On average, ... there was effectively no spin [z] component travelling along the path.*" Weak $B_z$ or strong $B_z$, there is absolutely no difference in the result of a spin $z$ measurement on either path. That is, you obtain $\langle \hat{\sigma}_z \rangle = 0$ not because "there was effectively no spin [z] component travelling along the path," but because the $z$ up and $z$ down outcomes occur with equal frequency, so they average to zero. Now we analyze what the weak values *do* mean in this experiment.

Per their Eq (1)

$$\langle \hat{A} \rangle_W = \frac{\langle \psi_f | \hat{A} | \psi_i \rangle}{\langle \psi_f | \psi_i \rangle}$$

per their Eq (2)

$$|\psi_i\rangle = \frac{1}{\sqrt{2}} \left( |S_x +\rangle |I\rangle + |S_x -\rangle |II\rangle \right)$$

and per their Eq (3)

$$|\psi_f\rangle = \frac{1}{\sqrt{2}}(|S_x -\rangle|I\rangle + |S_x -\rangle|II\rangle)$$

The Hilbert space is 4-dim, i.e., $|-\rangle|I\rangle$, $|+\rangle|I\rangle$, $|-\rangle|II\rangle$, and $|+\rangle|II\rangle$ (again, dropping the $S_x$ for brevity). $I^{REF} = |\langle\psi_f|\psi_i\rangle|^2 = \frac{1}{4}$ is the $\chi = 0$ intensity at O without an absorber or $B_z$ in the interferometer. We want to know how this $I^{REF}$ is affected by changes to the amplitude inside the interferometer that result from a weak absorber or weak $B_z$ in each path (four scenarios). Mathematically speaking, $\hat{A} = \hat{\Pi}_I$ in their Eq (1) gives $\langle\hat{\Pi}_I\rangle_W = 0$ because

$$\hat{\Pi}_I|\psi_i\rangle = \frac{1}{\sqrt{2}}|+\rangle|I\rangle \tag{5}$$

and $|\psi_f\rangle$ has no $|+\rangle|I\rangle$ component, since a $|-\rangle$ spin selector immediately precedes O. For $\hat{A} = \hat{\Pi}_{II}$, their Eq (1) gives $\langle\hat{\Pi}_{II}\rangle_W = 1$ because

$$\hat{\Pi}_{II}|\psi_i\rangle = \frac{1}{\sqrt{2}}|-\rangle|II\rangle \tag{6}$$

is the entirety of $|\psi_f\rangle$'s projection onto $|\psi_i\rangle$. These weak values are relevant to an absorber in either path because an absorber simply attenuates the amplitude at that point and that attenuation is propagated through to the amplitude at O. In other words, attenuating the $|+\rangle|I\rangle$ component of $|\psi_i\rangle$ has no effect on its projection on the final amplitude $|\psi_f\rangle$, so an absorber in path I has no effect on the intensity at O. However, attenuating the $|-\rangle|II\rangle$ part of $|\psi_i\rangle$ affects the entirety of $|\psi_i\rangle$'s projection on the final amplitude $|\psi_f\rangle$, so we expect this attenuation to be 100% transmitted to O (the second beam splitter is already taken into account in $|\psi_f\rangle$). Thus, their claim about how the weak values $\langle\hat{\Pi}_{II}\rangle_W = 1$ and $\langle\hat{\Pi}_I\rangle_W = 0$ bear on the intensity at O *for the absorber part of the experiment* is in accord with quantum mechanics. Again, there is no controversy for the absorber part of the experiment.

When $\hat{A} = \hat{\sigma}_z \hat{\Pi}_I$ we are projecting the $|+\rangle|I\rangle$ component of $|\psi_i\rangle$, then

$$\hat{\sigma}_z|+\rangle|I\rangle = -|-\rangle|I\rangle \tag{7}$$

is changing the $|+\rangle|I\rangle$ component of $|\psi_i\rangle$ that is perpendicular to $|\psi_f\rangle$ to $|\psi_i\rangle$'s entire projection on $|\psi_f\rangle$. Thus, $\langle \hat{\sigma}_z \hat{\Pi}_I \rangle_W = -1$. [Note: The sign discrepancy doesn't bear on the results, since this weak value is squared in the theoretical intensity, their Eq (10).] Likewise, when $\hat{A} = \hat{\sigma}_z \hat{\Pi}_{II}$ we are projecting the $|-\rangle|II\rangle$ component of $|\psi_i\rangle$, then

$$\hat{\sigma}_z|-\rangle|II\rangle = -|+\rangle|II\rangle \tag{8}$$

is changing this component so that it has no projection on $|\psi_f\rangle$, which yields $\langle \hat{\sigma}_z \hat{\Pi}_{II} \rangle_W = 0$. However, unlike $\langle \hat{\Pi}_I \rangle_W = 0$ in the absorber part of the experiment, the weak value $\langle \hat{\sigma}_z \hat{\Pi}_{II} \rangle_W = 0$ does not tell us that weak $B_z$ in path II has no effect on the intensity at O. The effect due to $B_z$ is given by the unitary operator (Denkmayr et al.'s Eq (8))

$$e^{i\hat{\sigma}_z \alpha/2} = \left( \hat{I} \cos\left(\frac{\alpha}{2}\right) + i\hat{\sigma}_z \sin\left(\frac{\alpha}{2}\right) \right)$$

acting on the amplitude at that point. Thus, the weak values $\langle \hat{\sigma}_z \hat{\Pi}_{II} \rangle_W$ and $\langle \hat{\sigma}_z \hat{\Pi}_I \rangle_W$ *are only giving us information about the effect of the sine term in* $e^{i\hat{\sigma}_z \alpha/2}$, i.e., the term that changes $|\pm\rangle$ into $|\mp\rangle$ in the amplitude at that point. So, the weak values $\langle \hat{\sigma}_z \hat{\Pi}_{II} \rangle_W = 0$ and $\langle \hat{\sigma}_z \hat{\Pi}_I \rangle_W = -1$ are telling us how much of the *change* $|\pm\rangle$ into $|\mp\rangle$ in $|\psi_i\rangle$ due to $B_z$ is propagated to O. When $B_z$ is placed in path I, 100% of this change in $|\psi_i\rangle$ matters at O because that change

$$\hat{\sigma}_z|+\rangle|I\rangle = -|-\rangle|I\rangle \tag{9}$$

is totally projected onto $|\psi_f\rangle$. When $B_z$ is placed in path II, none of this change in $|\psi_i\rangle$ matters at O because that change

$$\hat{\sigma}_z|-\rangle|II\rangle = -|+\rangle|II\rangle \tag{10}$$

has no projection onto $|\psi_f\rangle$ (again, because of the $|-\rangle$ spin selector immediately preceding O). In order to account for the *entire effect* of $e^{i\hat{\sigma}_z \alpha/2}$ on $|\psi_i\rangle$, and therefore the intensity at O, *we must also account for the effect of the cosine term in* $e^{i\hat{\sigma}_z \alpha/2}$ and that effect is accounted for by $\langle \hat{\Pi}_{II} \rangle_W = 1$ in their Eq (10) $I_j^{MAG} = |\langle \psi_f | \psi_i \rangle|^2 \left[ 1 - \frac{\alpha^2}{4} \langle \hat{\Pi}_j \rangle_W + \frac{\alpha^2}{4} |\langle \hat{\sigma}_z \hat{\Pi}_j \rangle_W|^2 \right]$ which gives our Eq (1), $I_{II/O}^{MAG} \approx I^{REF} \left( 1 - \left( \frac{\alpha}{2} \right)^2 \right)$. Thus, the weak value $\langle \hat{\Pi}_{II} \rangle_W = 1$ is simply telling us that *any* attenuation or growth in the $|-\rangle|II\rangle$ part of the amplitude will have an effect at O. [In their Eq (6), the absorber in path II introduces an attenuation in this part of the amplitude, so $\langle \hat{\Pi}_{II} \rangle_W = 1$ appears there as well.] The amount of that effect for weak $B_z$ is a loss of $\left( \frac{\alpha}{2} \right)^2$ coming from the (small α) action of $\hat{I}\cos\left(\frac{\alpha}{2}\right)$ in $e^{i\hat{\sigma}_z \alpha/2}$ on the amplitude in path II.

$\langle \hat{\sigma}_z \hat{\Pi}_{II} \rangle_W = 0$ is simply telling us that there is no contribution to the intensity at O due to the action of $i\hat{\sigma}_z \sin\left(\frac{\alpha}{2}\right)$ in $e^{i\hat{\sigma}_z \alpha/2}$ on the amplitude in path II. So, the explanatory role of $\langle \hat{\Pi}_{II} \rangle_W = 1$ is not limited to the absorber part of the experiment, it also plays a role in explaining the effect of weak $B_z$ in path II. *Therefore, we cannot infer that* $\langle \hat{\sigma}_z \hat{\Pi}_{II} \rangle_W = 0$ *means weak $B_z$ has no effect in path II, as is crucial to the quantum Cheshire Cat interpretation*. To understand the effect of weak $B_z$ in path II, you also have to take into account the (small α) action of $\hat{I}\cos\left(\frac{\alpha}{2}\right)$ in $e^{i\hat{\sigma}_z \alpha/2}$ on the amplitude in path II, and the relevant weak value for that is $\langle \hat{\Pi}_{II} \rangle_W = 1$. Continuing, $\langle \hat{\sigma}_z \hat{\Pi}_I \rangle_W = -1$ in $I_j^{MAG} = |\langle \psi_f | \psi_i \rangle|^2 \left[ 1 - \frac{\alpha^2}{4} \langle \hat{\Pi}_j \rangle_W + \frac{\alpha^2}{4} |\langle \hat{\sigma}_z \hat{\Pi}_j \rangle_W|^2 \right]$, which gives our Eq (2), $I_{I/O}^{MAG} \approx I^{REF} \left( 1 + \left( \frac{\alpha}{2} \right)^2 \right)$, is simply telling us that there is a gain of $\left( \frac{\alpha}{2} \right)^2$ in the intensity at O caused by weak $B_z$ coming from the (small α) action of $i\hat{\sigma}_z \sin\left(\frac{\alpha}{2}\right)$ in $e^{i\hat{\sigma}_z \alpha/2}$ on the amplitude in path I. $\langle \hat{\Pi}_I \rangle_W = 0$ is simply telling us that the action of $\hat{I}\cos\left(\frac{\alpha}{2}\right)$ in $e^{i\hat{\sigma}_z \alpha/2}$ on the amplitude in path I has no effect on the intensity at O. So, as with $\langle \hat{\Pi}_{II} \rangle_W = 1$ and $\langle \hat{\sigma}_z \hat{\Pi}_{II} \rangle_W = 0$ for path II, both $\langle \hat{\Pi}_I \rangle_W = 0$ and $\langle \hat{\sigma}_z \hat{\Pi}_I \rangle_W = -1$ are needed to describe the effect of weak $B_z$ in path I. We now show why this means it is impossible to conclude empirically, per weak values or otherwise, that "*the neutron and [the z component of] its spin are spatially separated*" in this experiment.

## 5. QUADRATIC INTERACTION DESTROYS QUANTUM CHESHIRE CAT

So, as we stated at the outset, per their claim, "Therefore, any probe system that interacts with the Cheshire Cat system *weakly enough* will on average be affected as if the neutron and [the *z* component of] its spin are spatially separated [our emphasis]," they needed to show that it is possible to introduce a weak enough $B_z$ on path II such that there would be no evidence of spin coupling on that path, while that same $B_z$ on path I would have an observable effect. To explore this possibility, consider the effect of $e^{i\hat{\sigma}_z \alpha/2}$ on the amplitude in path *j* to lowest (linear) order in α, i.e.,

$$e^{i\hat{\sigma}_z \alpha/2} \approx \hat{I} + i\hat{\sigma}_z \alpha / 2 \qquad (11)$$

Accordingly, when weak $B_z$ is placed in path *j*, all that matters to lowest order in α for understanding the effect *on the amplitude* is $i\hat{\sigma}_z \alpha / 2$, which is characterized by $\langle \hat{\sigma}_z \hat{\Pi}_j \rangle_W$. Notice that by "weak" we mean "weak enough" to exclude measureable/observable second-order (quadratic) effects in α. So, with weak $B_z$ in path I, given the effect of $\hat{\sigma}_z$ on $|S_x +\rangle$ and the $|S_x -\rangle$ spin selector immediately preceding detector O, there is a discernible effect of the interaction Hamiltonian *on the amplitude* and $\langle \hat{\sigma}_z \hat{\Pi}_I \rangle_W = -1$ captures this fact nicely. In contrast, with weak $B_z$ in path II, the effect of the interaction Hamiltonian on the amplitude is given by $\hat{I}$, i.e., it has no effect *on the amplitude*, and $\langle \hat{\sigma}_z \hat{\Pi}_{II} \rangle_W = 0$ captures this fact nicely. So, as we posited, "weak enough" means "linear interaction." If this is possible, then of course we should see no effect on the intensity at O, i.e., no change in $I^{REF}$, when introducing this weak $B_z$ to path II, since all we have introduced to the amplitude at O is $\hat{I}$. If that happened, the authors would certainly be justified in saying, "It's as if there is no spin coupling on path II for weak $B_z$," i.e., we could infer the quantum Cheshire Cat interpretation from the weak values. But, as we saw, this isn't what happens at O.

Rather, as we showed above, there is a reduction in the intensity at O for weak $B_z$ in path II no matter how weak we make $B_z$. The reason for this is that the intensity is obtained from the amplitude squared, not the amplitude. So, when we want to ask a question about the *empirical* effect of the interaction Hamiltonian *on the amplitude* for weak $B_z$ in path II, we must keep the quadratic term in α coming from $e^{i\hat{\sigma}_z\alpha/2}$, because ultimately that term will be as important observationally, i.e., *for the intensity*, as the linear term in α coming from $e^{i\hat{\sigma}_z\alpha/2}$ *no matter how weak we make $B_z$*. That is,

$$e^{i\hat{\sigma}_z\alpha/2} \approx \hat{I} + i\hat{\sigma}_z\alpha/2 - \hat{I}\alpha^2/8 = \hat{I}\left(1-\alpha^2/8\right) + i\hat{\sigma}_z\alpha/2 \qquad (12)$$

is the form that gives $I_j^{MAG} = \left|\langle\psi_f|\psi_i\rangle\right|^2 \left[1 - \frac{\alpha^2}{4}\langle\hat{\Pi}_j\rangle_W + \frac{\alpha^2}{4}\left|\langle\hat{\sigma}_z\hat{\Pi}_j\rangle_W\right|^2\right]$. So, the term $\frac{\alpha^2}{4}\langle\hat{\Pi}_j\rangle_W$ in $I_j^{MAG}$ that destroys the quantum Cheshire Cat interpretation comes from the quadratic term in the expansion of $e^{i\hat{\sigma}_z\alpha/2}$. Thus, the quadratic piece of the $B_z$ interaction cannot be avoided and it entails just the opposite of the quantum Cheshire Cat interpretation, i.e., it means that, "No matter how weak we make $B_z$, there will be empirical evidence for a spin *z* coupling on path II."

## 6. CONCLUSION

In conclusion, Denkmayr *et al.* have a beautiful marriage of experiment with simple quantum mechanics theory. However, while the weak values in their experiment accurately accounted for the measured neutron intensities, they do not show, entail, or in any way suggest that a particle and one of its properties have been spatially separated, i.e., they do not imply quantum Cheshire Cat. This is because the weak values were measured with an unavoidable quadratic $B_z$ interaction and the quadratic $B_z$ interaction means there will be empirical evidence for a spin *z* coupling on both paths of the interferometer, no matter how weak you make $B_z$, which renders the quantum Cheshire Cat interpretation untenable. While Denkmayr *et al.* failed to instantiate quantum

Cheshire Cat, their experiment does reveal an interesting fact, i.e., weak values can be measured via quadratic interaction, but weak values measured in this fashion may not have straightforward ontological implications. Thus, their experiment should prompt further discussion in the weak values community.

**Denkmayr *et al*. Figure 1
Quantum Cheshire Cat Experiment**

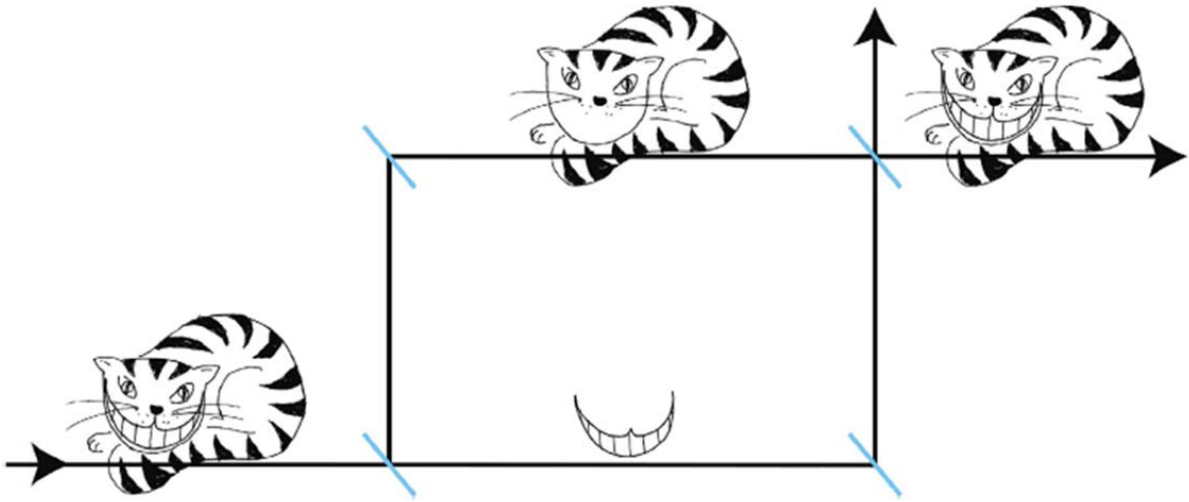

**Figure 2
Result of Denkmayr *et al*. Experiment**

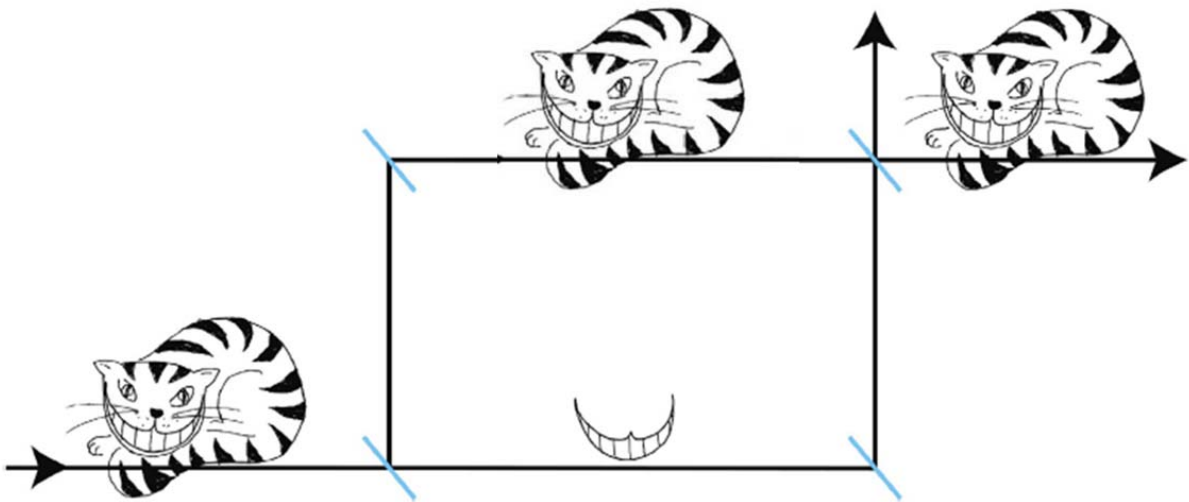

**Denkmayr *et al*. Figure 3**
**Absorber**

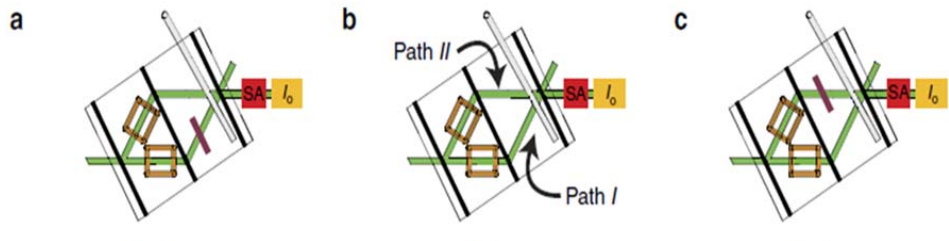

**Denkmayr *et al*. Figure 4**
**Magnetic Field**

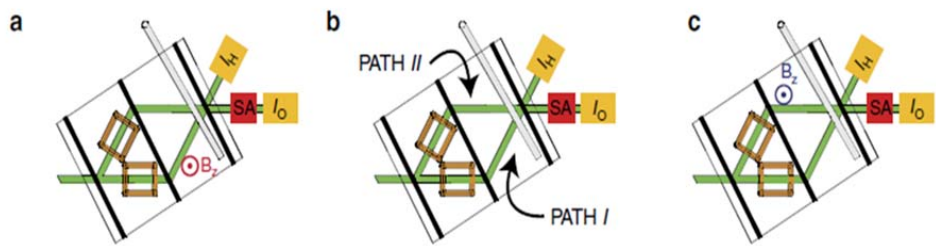